\documentclass[journal]{IEEEtran}
\usepackage{amsmath,amsfonts}
\usepackage{algorithmic}
\usepackage{algorithm}
\usepackage{array}
\usepackage[caption=false,labelfont=rm,textfont=rm]{subfig}
\usepackage{textcomp}
\usepackage{stfloats}
\usepackage{url}
\usepackage{verbatim}
\usepackage{graphicx}
\usepackage{cite}
\usepackage{amsmath,amsfonts}
\usepackage{amssymb}
\usepackage{mathtools}
\usepackage{mathrsfs}
\usepackage{bm}
\usepackage{braket}
\usepackage{algorithmic}
\usepackage{algorithm}
\usepackage{array}
\usepackage{microtype}
\usepackage{textcomp}
\usepackage{stfloats}
\usepackage{url}
\usepackage{verbatim}
\usepackage{graphicx}
\usepackage{cite}
\usepackage{xcolor}
\usepackage{xfrac}
\usepackage[utf8]{inputenc}
\usepackage[T1]{fontenc}
\usepackage{balance}
\usepackage{multirow}
\usepackage{multicol}
\usepackage{booktabs}
\usepackage{setspace}
\usepackage{optidef}
\def\BibTeX{{\rm B\kern-.05em{\sc i\kern-.025em b}\kern-.08em
    T\kern-.1667em\lower.7ex\hbox{E}\kern-.125emX}}

\DeclareMathOperator*{\sign}{sign}
\DeclareMathOperator*{\diag}{diag}
\DeclareMathOperator*{\R}{R}

\DeclareMathOperator*{\bdiag}{bdiag}

\begin{document}

\title{Quasi-Newton FDE in One-Bit Pseudo-Randomly Quantized Massive MIMO-OFDM Systems}

\author{Gökhan Yılmaz, \IEEEmembership{Graduate Student Member, IEEE}, and Ali Özgür Yılmaz, \IEEEmembership{Member, IEEE}
\thanks{G. Yılmaz (g.yilmaz@tue.nl) was with the Department of Electrical and Electronics Engineering, Middle East Technical University, Ankara, Turkey, during this work. He is now with the Department of Electrical Engineering, Eindhoven University of Technology, Eindhoven, The Netherlands.}
\thanks{A. Ö. Yılmaz (aoyilmaz@metu.edu.tr) is with the Department of Electrical and Electronics Engineering, Middle East Technical University, Ankara, Turkey.}
\thanks{The work of G. Yılmaz was supported by Vodafone Turkey within the framework of the 5G and Beyond Joint Graduate Support Program coordinated by the Information and Communication Technologies Authority of Turkey.}
}
% Remember, if you use this, you must call \IEEEpubidadjcol in the second

\maketitle

\begin{abstract}
This letter offers a new frequency-domain equalization (FDE) scheme that can work with a pseudo-random quantization (PRQ) scheme utilizing non-zero threshold quantization in one-bit uplink multi-user massive multiple-input multiple-output (MIMO) systems to mitigate quantization distortion and support high-order modulation schemes. The equalizer is based on Newton's method (NM) and applicable for orthogonal frequency division multiplexing (OFDM) transmission under frequency-selective fading by exploiting the properties of massive MIMO. We develop a low-complexity FDE scheme to obtain a quasi-Newton method. The proposed detector outperforms the benchmark detector with comparable complexity.
\end{abstract}


\section{Introduction}

\IEEEPARstart{M}{assive} multiple-input multiple-output (MIMO) systems are equipped with large numbers of antennas to achieve higher data rates. However, cost and power consumption can be significant issues since each antenna requires a separate radio frequency (RF) chain. Low-resolution analog-to-digital converters (ADCs) have been proposed as a potential solution \cite{1BOX, BayesOptimal_OFDM, Wideband_1bit_Perf, EM_OFDM, IEM_OFDM, oversampling_perf} since ADCs are one of the most power consuming components in the RF chains. The power consumption of an ADC increases exponentially with its resolution. One-bit ADCs have additional benefits, such as not requiring automatic gain control units and having simple circuitry. However, communication with high-order QAM constellations is challenging to achieve with one-bit ADCs due to the loss of amplitude information. The achievable rate of one-bit systems saturate at a high signal-to-noise ratio (SNR) value due to the stochastic resonance (SR) phenomenon \cite{OBNHD}.

In our previous work \cite{OBNHD}, we aimed to tackle this issue by proposing new quantization and detection schemes for one-bit massive MIMO systems operating under frequency-flat fading. Inspired by the idea of randomizing the quantization operation to lower the quantization distortion in the temporal domain for general signal processing application from \cite{DAFSP_PRQ}, we introduced a new pseudo-random quantization (PRQ) scheme by changing the domain of dithering from temporal to spatial using quantization thresholds, in contrast to existing works such as \cite{1BOX, BayesOptimal_OFDM, Wideband_1bit_Perf, EM_OFDM, IEM_OFDM, oversampling_perf} where the quantization thresholds are zero for all receiver antennas i.e., zero-threshold quantization (ZTQ) is employed. In \cite{OBNHD}, we show that by employing the PRQ approach, one-bit uplink massive MIMO systems can operate with high-order QAM constellations such as $256$-QAM or $1024$-QAM by compensating for the SR effects.

In this letter, we extend our previous work \cite{OBNHD} to the frequency-selective fading scenario with orthogonal frequency division multiplexing (OFDM) to analyze the effects of PRQ when there is inter-symbol interference (ISI) on the received signal. We formulate equalization in one-bit wideband massive MIMO systems as a constrained optimization problem by optimizing over the log-likelihood function. Unlike \cite{1BOX}, this approach does not require selecting suitable step sizes. After deriving the necessary relations using Newton's method (NM), we utilize two approximations based on the properties of the Hessian of the log-likelihood and massive MIMO to decouple equalization among subcarriers and avoid matrix inversion to reach the proposed low-complexity frequency-domain equalization (FDE) scheme, projected quasi-Newton detector (PQND), which is an improvement over \cite{OBNHD}. In one-bit massive MIMO systems, adopting second-order optimization techniques to optimize the log-likelihood under frequency-selective fading has not been used. The proposed detector outperforms the benchmark detector from \cite{1BOX}, irrespective of the quantization scheme with comparable complexity. We also show that communication with high-order modulation schemes is applicable when PRQ is used in frequency-selective channels.

\textit{Notation:} Lowercase letters represent scalars, lowercase bold letters represent vectors, and uppercase bold letters represent matrices. Underline accent is used to represent the concatenated version of a time-domain (TD) vector over all $V$ time samples or a frequency-domain (FD) vector over all $V$ subcarriers such that $\bar{\underline{\boldsymbol{a}}} = \begin{bmatrix} \bar{\boldsymbol{a}}[0]^T & \bar{\boldsymbol{a}}[1]^T & \hdots & \bar{\boldsymbol{a}}[V-1]^T \end{bmatrix} ^T.$ Complex-valued variables have an over-bar accent and their real-valued counterparts have no accent on top such that

\begin{equation}
    \label{eqn:vector_real}
    \underline{\boldsymbol{a}} =
    \begin{bmatrix}
    \Re\{\bar{\underline{\boldsymbol{a}}}\} \\ \Im\{\bar{\underline{\boldsymbol{a}}}\}
    \end{bmatrix} \textrm{ and }
    \underline{\boldsymbol{A}} =
    \begin{bmatrix}
    \Re\{\bar{\underline{\boldsymbol{A}}}\} & -\Im\{\bar{\underline{\boldsymbol{A}}}\} \\
    \Im\{\bar{\underline{\boldsymbol{A}}}\} & \Re\{\bar{\underline{\boldsymbol{A}}}\}
    \end{bmatrix}.
\end{equation}

\noindent $\bar{\boldsymbol{F}}$ is the unitary discrete Fourier transform (DFT) matrix of size $V \times V$. $\bdiag \left[.\right]$ operator creates a block diagonal matrix by placing its argument entries on the diagonal as blocks. $\otimes$ is the Kronecker and $\odot$ is the Hadamard product. A variation $\bar{\odot}$ of the Hadamard product is defined as $\bar{\boldsymbol{a}} \, \bar{\odot} \, \bar{\boldsymbol{b}} = \Re\{\bar{\boldsymbol{a}}\} \odot \Re\{\bar{\boldsymbol{b}}\} + j \, \Im\{\bar{\boldsymbol{a}}\} \odot \Im\{\bar{\boldsymbol{b}}\}$.

\section{System Model} \label{sec:system_model}

\begin{figure}[t]
    \centering
    \includegraphics[width=\columnwidth]{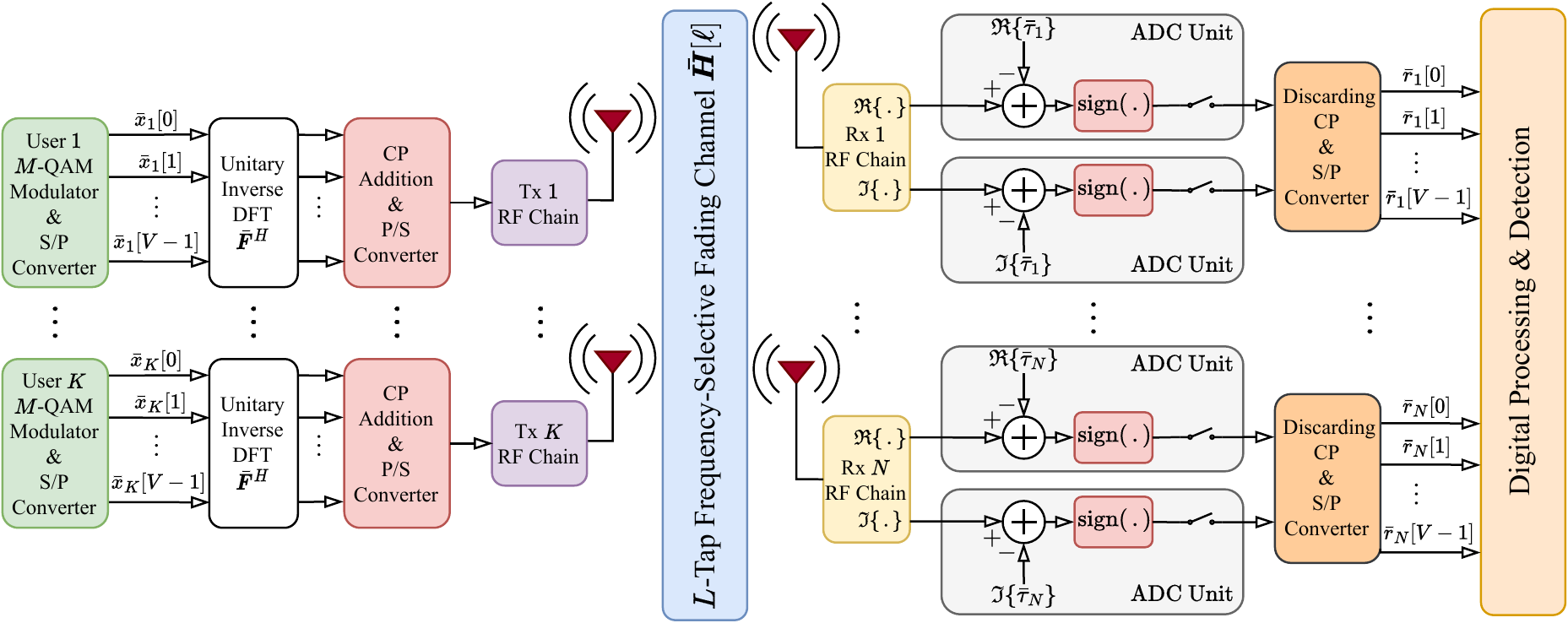}
    \caption{A block diagram that summarizes the system model. S/P is for serial-to-parallel, and P/S is for parallel-to-serial.}
    \label{fig:system_model}
\end{figure}

We consider an $N \times K$ uplink massive MIMO-OFDM system with $V$ subcarriers, where $K$ single-antenna users are served by a base station (BS) equipped with $N$ antennas. A summary of the system model is shown in Fig. \ref{fig:system_model}. Constellation symbols to be transmitted are selected independently from an $M$-QAM alphabet, $\bar{\mathcal{M}}$, with equal likelihood. The FD symbols of the users at the $v^{\textrm{th}}$ subcarrier are shown as $\bar{\boldsymbol{x}}[v] = \begin{bmatrix} \bar{x}_1[v] & \bar{x}_2[v] & \hdots & \bar{x}_K[v] \end{bmatrix}^T \in \bar{\mathcal{M}}^K$ for $v=0,1,\hdots, V-1$, where $\mathbb{E}[|\bar{x}_{k}[v]|^2]=E_s=1$. Each user transmits their FD symbols through inverse DFT before transmission to obtain their TD signals. We assume the BS has perfect knowledge of the channel impulse response (CIR), and the multipath channel has $L$ taps. Each user adds a CP of length $L_{\textrm{CP}} \geq L-1$ to the beginning of their TD signals to help mitigate the effects of ISI at the receiver side. The $\ell^{\textrm{th}}$ tap of CIR between all users and BS antennas can be represented as $\bar{\boldsymbol{H}}[\ell] \in \mathbb{C}^{N \times K}$, where the channel between the $n^{\mathrm{th}}$ BS antenna and the $k^{\mathrm{th}}$ user can be shown as $\Bar{h}_{n,k}[\ell]$ with $\bar{{h}}_{n,k}[\ell] = 0$ for $\ell \geq L$. The CP is discarded at the receiver side. $\Bar{\boldsymbol{y}}[m] = \begin{bmatrix} \bar{y}_1[m] & \bar{y}_2[m] & \hdots & \bar{y}_N[m] \end{bmatrix}^T \in \mathbb{C}^N$ and $\Bar{\boldsymbol{w}}[m] = \begin{bmatrix} \bar{w}_1[m] & \bar{w}_2[m] & \hdots & \bar{w}_N[m] \end{bmatrix}^T \in \mathbb{C}^N$ represent the unquantized version of the received signal and the noise samples time $m$, respectively. $\bar{\boldsymbol{w}}[m] \sim \mathcal{CN}(\boldsymbol{0}_N, N_0 \boldsymbol{I}_N)$ for $m=0,1,\hdots, V-1$. We define the TD block circulant MIMO channel matrix of size $NV \times KV$ as

\begin{equation}
    \label{eqn:circulant_channel_matrix}
    \begin{split}
        \underline{\bar{\boldsymbol{H}}}_b & = 
        \begin{bmatrix}
            \bar{\boldsymbol{H}}[0] & \bar{\boldsymbol{H}}[V-1] & \hdots & \bar{\boldsymbol{H}}[1] \\
            \bar{\boldsymbol{H}}[1] & \bar{\boldsymbol{H}}[0] & \hdots & \bar{\boldsymbol{H}}[2] \\
            \vdots & \ddots & \ddots & \vdots \\
            \bar{\boldsymbol{H}}[V-1] & \hdots & \bar{\boldsymbol{H}}[1] & \bar{\boldsymbol{H}}[0] 
            \end{bmatrix}
            \\
            & = \bar{\boldsymbol{Q}}_N^H \underline{\bar{\boldsymbol{\Lambda}}}_b \bar{\boldsymbol{Q}}_K,
    \end{split}
\end{equation}

\noindent where $\bar{\boldsymbol{Q}}_i = \bar{\boldsymbol{F}} \otimes \boldsymbol{I}_i$ for $i \in \mathbb{Z}^+$. $\underline{\bar{\boldsymbol{\Lambda}}}_b \in \mathbb{C}^{NV \times KV}$ is the block diagonal FD MIMO channel matrix \cite{oversampling_ofdm}. It can be shown as $\underline{\bar{\boldsymbol{\Lambda}}}_b = \bdiag \left[ \bar{\boldsymbol{\Lambda}}[0], \bar{\boldsymbol{\Lambda}}[1], \hdots, \bar{\boldsymbol{\Lambda}}[V-1] \right]$. The TD unquantized representation of the received signal can be found as

\begin{equation}
    \label{eqn:complex_io_unq}
    \begin{split}
        \underline{\bar{\boldsymbol{y}}}
        & = \underline{\bar{\boldsymbol{H}}}_b \, \bar{\boldsymbol{Q}}_K^H \, \underline{\bar{\boldsymbol{x}}} + \underline{\bar{\boldsymbol{w}}} \\
        & = \underline{\bar{\boldsymbol{G}}} \, \underline{\bar{\boldsymbol{x}}} + \underline{\bar{\boldsymbol{w}}},
    \end{split}
\end{equation}

\noindent where $\underline{\bar{\boldsymbol{G}}} = \bar{\boldsymbol{Q}}_N^H \underline{\bar{\boldsymbol{\Lambda}}}_b$ represents the effective TD channel matrix. $\underline{\bar{\boldsymbol{w}}} \in \mathbb{C}^{NV}$ is the TD concatenated noise vector with independent entries. Then, the concatenated representation $\underline{\bar{\boldsymbol{r}}}$ of the TD one-bit quantized observations $\Bar{\boldsymbol{r}}[m] = \begin{bmatrix} \bar{r}_1[m] & \bar{r}_2[m] & \hdots & \bar{r}_N[m] \end{bmatrix}^T \in \{ \pm 1 \pm j\}^N$ is found as

\begin{equation}
    \label{eqn:complex_io_quan}
    \underline{\bar{\boldsymbol{r}}} = \sign(\Re\{\underline{\bar{\boldsymbol{y}}} - \underline{\bar{\boldsymbol{\tau}}}\}) + j \, \sign(\Im\{\underline{\bar{\boldsymbol{y}}} - \underline{\bar{\boldsymbol{\tau}}}\}),
\end{equation}

\noindent where $\underline{\bar{\boldsymbol{\tau}}}$ is the time-invariant quantization threshold vector with components $\bar{\boldsymbol{\tau}}[m_1]=\bar{\boldsymbol{\tau}}[m_2]=\bar{\boldsymbol{\tau}}$ for $m_1, m_2 \in \{0,1,\hdots,V-1\}$. The quantization thresholds are time-invariant. Due to the structure of the proposed detector, we need to be able to work with real numbers only. Hence, (\ref{eqn:complex_io_quan}) can be re-written as $\underline{\boldsymbol{r}} = \sign \left( \underline{\boldsymbol{G}} \, \underline{\boldsymbol{x}} - \underline{\boldsymbol{\tau}} + \underline{\boldsymbol{w}} \right)$. The channel is modeled with Rayleigh fading, i.e., $\bar{h}_{n,k}[\ell] \sim \mathcal{CN}(0,p[\ell])$, where $p[\ell]$ denotes the power delay profile (PDP) of the channel and each coefficient is independent in space and time. The power of each tap in an exponential PDP channel can be calculated as $p[\ell] = \frac{\exp(-\mu \ell)}{\sum_{m=0}^{L-1} \exp(-\mu m)}$ where $\mu$ is the power decay rate. To model a small delay spread (SDS) channel, we use exponential PDP where $L=8$ and $\mu=1$. To model a large delay spread (LDS) channel, we utilize TDL-A delay profile from \cite{5g_tdl}, a tapped delay line channel model introduced by 3GPP, for which $L=23$. The sum power from all taps in both models is $\sum_{\ell=0}^{L-1} p[\ell] = 1$. We define the SNR as the ratio of the average received signal power per user to the average noise power at each antenna such that $\rho=\sfrac{E_s}{N_0}=\sfrac{1}{N_0}$. 
 
\section{Projected Quasi-Newton Detector (PQND)} \label{sec:obqnd}

In the presence of independent additive white Gaussian noise (AWGN) on each branch of a one-bit quantized channel, as in \cite{1BOX, OBNHD}, the log-likelihood can be expressed as

\begin{equation}
    \label{eqn:log_likelihood}
    \mathcal{L}(\underline{\boldsymbol{x}})
    = \boldsymbol{1}_{2NV}^T \log \left( \Phi \left( \sqrt{\frac{2}{N_0}} \underline{\boldsymbol{r}} \odot ( \underline{\boldsymbol{G}} \, \underline{\boldsymbol{x}} - \underline{\boldsymbol{\tau}}) \right) \right),
\end{equation}

\noindent where $\Phi(a) = \int_{-\infty}^{a} \phi(\tau) \, d\tau$ is the cumulative distribution function (CDF) of the standard Gaussian random variable with probability density function (PDF) $\phi(a) = \frac{\exp(-a^2)}{\sqrt{2 \pi}}$, and $\boldsymbol{1}_i$ is the vector of size $i \in \mathbb{Z}^+$, whose elements are all set to $1$. $\log(.)$ and $\Phi(.)$ are applied element-wise on their arguments. (\ref{eqn:log_likelihood}) calculates the logarithm of the probability of observing $\underline{\boldsymbol{r}}$ when $\underline{\boldsymbol{x}}$ passes through the channel $\underline{\boldsymbol{G}}$ and gets one-bit quantized with the thresholds $\underline{\boldsymbol{\tau}}$. For a single branch ADC with a $+1$ observation, (\ref{eqn:log_likelihood}) is the logarithm of the probability that the noise-free version of the received signal is greater than the threshold, given the transmitted signal. The optimal solution that maximizes (\ref{eqn:log_likelihood}) requires an exhaustive search. Subcarrier-level maximum likelihood (ML) detection in the FD is not applicable since conversion to the FD causes an unknown conditional distribution for the quantized observations. We use gradient-based optimization techniques \cite{OBNHD, 1BOX} by utilizing the concavity of (\ref{eqn:log_likelihood}) due to the log-concavity of $\Phi(.)$. We relax the discrete input set constraint to get

\begin{argmaxi}|l|[2]
{\scriptstyle{\underline{\boldsymbol{x}}} \in \mathbb{R}^{2KV}}
{\mathcal{L}(\underline{\boldsymbol{x}})}
{\label{eqn:opt_problem}}{\underline{\tilde{\boldsymbol{x}}}=}
\addConstraint{|\underline{x}_i|}{\leq M_b, \quad}{i=1,2,\hdots,2KV},
\end{argmaxi}

\noindent where $M_b$ is the boundary of an $M$-QAM constellation with unit average power on a single dimension, and it can be calculated as $M_b = \sqrt{\frac{3 (\sqrt{M}-1)^2}{2 (M-1)}}$. The feasibility of relaxation and the convergence of concave optimization problems of similar forms is shown in \cite{alg_convergence}. While relaxing the discrete input set constraint, we utilize a restriction. Each element of $\underline{\boldsymbol{x}}$ is inside the boundaries of the constellation, which is called the box constraint \cite{1BOX}. The box constraint prevents diverging from the constellation's boundaries during equalization. To apply constrained optimization, we resort to the projection method. We start with projected NM, a second-order optimization technique, to solve (\ref{eqn:opt_problem}). The update equation can be written as

\begin{equation}
    \label{eqn:newtons_update}
    \underline{\boldsymbol{x}}^{(t)}
    = \mathcal{P}^{(t)} \left( \underline{\boldsymbol{x}}^{(t-1)} - \alpha (\nabla^2 \mathcal{L}^{(t-1)})^{-1} \nabla \mathcal{L}^{(t-1)} \right),
\end{equation}

\noindent where $\alpha \in \mathbb{R}$ is the step size, and $\mathcal{P}^{(t)}(.)$ is the projection function at iteration $t$, which will be addressed after the gradient and Hessian calculations. $\nabla \mathcal{L}^{(t)}$ is the gradient of the log-likelihood function $\mathcal{L}$ with respect to $\underline{\boldsymbol{x}}$ calculated at iteration $t$. Similarly, $\nabla^2 \mathcal{L}^{(t)}$ is the Hessian of $\mathcal{L}$ with respect to $\underline{\boldsymbol{x}}$ at iteration $t$. The gradient and Hessian are functions of $\underline{\boldsymbol{x}}$, and we drop the argument for ease of notation. Also, we define $\underline{\boldsymbol{u}} = \sqrt{\frac{2}{N_0}} \underline{\boldsymbol{r}} \odot ( \underline{\boldsymbol{G}} \, \underline{\boldsymbol{x}} - \underline{\boldsymbol{\tau}})$ for compactness. The gradient is

\begin{equation}
\label{eqn:gradient}
    \nabla \mathcal{L}
    = \sqrt{\frac{2}{N_0}} \underline{\boldsymbol{G}}^T \left( \underline{\boldsymbol{r}} \odot \varphi \left( \underline{\boldsymbol{u}} \right) \right),
\end{equation}

\noindent where $\varphi(a) \triangleq \frac{d}{da} \ln(\Phi(a)) = \phi(a) / \Phi(a) > 0$ since both PDF and CDF are non-negative functions, and $\varphi(.)$ is applied element-wise on its arguments. The Hessian can be found as

\begin{equation}
\label{eqn:hessian}
    \nabla ^2 \mathcal{L}
    = \frac{2}{N_0} \underline{\boldsymbol{G}}^T \diag \left( \psi \left( \underline{\boldsymbol{u}} \right) \right) \underline{\boldsymbol{G}},
\end{equation}

\noindent where $\psi(a) \triangleq \frac{d^2}{da^2} \ln(\Phi(a)) = -a\varphi(a) -\varphi^2(a) < 0$ since $\varphi(a)>0$, and $\psi(.)$ is applied element-wise on its arguments. Note that $\psi(a) < 0$ for all $a \in \mathbb{R}$, which shows the log-concavity of $\Phi(.)$. The computations of these nonlinear functions in finite precision can cause divergent behavior for which the solution in \cite[Appendix~B]{OBNHD} is also used herein. NM can provide complexity reduction. However, it does not allow subcarrier-level equalization and requires a large matrix inversion. We solve these problems using approximations to get a low-complexity equalizer. We can write the step to be computed in (\ref{eqn:step}), $\Delta \underline{\boldsymbol{x}} = (\nabla ^2 \mathcal{L})^{-1} \nabla \mathcal{L}$, as

\begin{equation}
    \label{eqn:step}
        \Delta \underline{\boldsymbol{x}}
        = \sqrt{\frac{N_0}{2}} \left( \underline{\boldsymbol{G}}^T \diag \left( \psi \left( \underline{\boldsymbol{u}} \right) \right) \underline{\boldsymbol{G}} \right) ^{-1} \left( \underline{\boldsymbol{G}}^T \left( \underline{\boldsymbol{r}} \odot \varphi \left( \underline{\boldsymbol{u}} \right) \right) \right).
\end{equation}

The matrix to be inverted in (\ref{eqn:step}) is a $2KV \times 2KV$ matrix. To simplify the relations, we first aim to approximate $\diag(\psi(\underline{\boldsymbol{u}}))$ to a multiple of the identity matrix $\gamma \boldsymbol{I}_{2NV}$ such that $\gamma = \frac{1}{2NV} \sum_{n=1}^{2NV} \psi(\underline{u}_n)$. This approximation is valid for the low SNR regime when $N_0$ is large so that the elements of $\underline{\Bar{\boldsymbol{u}}}$ are very close to zero. For the high SNR regime, it is very likely that all elements of $\underline{\Bar{\boldsymbol{u}}}$ are positive and large values. Since the rate of change of $\psi(.)$ is very low for large positive arguments, this approximation is also helpful for high SNR. Note that with this approximation, decoupling between the real and imaginary parts due to second-order derivative calculation can also be avoided. We can now go back to the notation with complex numbers. The complex counterpart of (\ref{eqn:newtons_update}) can be obtained as $\underline{\bar{\boldsymbol{x}}}^{(t)} = \mathcal{P}^{(t)} \left( \underline{\bar{\boldsymbol{x}}}^{(t-1)} - \alpha \Delta \underline{\bar{\boldsymbol{x}}}^{(t-1)} \right)$, where $\Delta \underline{\bar{\boldsymbol{x}}}$ is found as

\begin{equation}
    \label{eqn:complex_step}
    \Delta \underline{\bar{\boldsymbol{x}}} \cong
    \frac{1}{\gamma} \sqrt{\frac{N_0}{2}} \left( \bar{\underline{\boldsymbol{\Lambda}}}_b^H \bar{\underline{\boldsymbol{\Lambda}}}_b \right)^{-1} \bar{\underline{\boldsymbol{\Lambda}}}_b^H \bar{\boldsymbol{Q}}_N \left( \bar{\underline{\boldsymbol{r}}} \, \bar{\odot} \, \bar{\varphi} \left( \bar{\underline{\boldsymbol{u}}} \right) \right),
\end{equation}

\noindent by defining $\bar{\varphi}(\bar{\boldsymbol{a}}) = \varphi(\Re\{\bar{\boldsymbol{a}}\}) + j \, \varphi(\Im\{\bar{\boldsymbol{a}}\})$. Note that $\underline{\bar{\boldsymbol{u}}}$ from its each component $\{ \bar{\boldsymbol{u}}[m] \}_{m=0}^{V-1}$ can now be found as 

\begin{equation}
    \label{eqn:complex_arg_vector}
    \bar{\boldsymbol{u}}[m] = \sqrt{\frac{2}{N_0}} \bar{\boldsymbol{r}}[m] \, \bar{\odot} \, \left( \mathcal{F}_m^{-1} \{ \bar{\boldsymbol{\Lambda}}[v] \bar{\boldsymbol{x}}[v] \} - \bar{\boldsymbol{\tau}} \right),
\end{equation}

\noindent where $\mathcal{F}_m^{-1}(.) = \sum_{v=0}^{V-1} (.) \exp \left(j 2 \pi \frac{mv}{V}\right) $. Since $\underline{\bar{\boldsymbol{\Lambda}}}_b$ is a block diagonal matrix, subcarrier-level processing is now applicable. (\ref{eqn:complex_step}) can be written for each subcarrier $v=0,1,\hdots,V-1$ as 

\begin{equation}
    \label{subeqn:step1_sc}
    \Delta \bar{\boldsymbol{x}}[v] \cong \frac{1}{ \gamma} \sqrt{\frac{N_0}{2}} \bar{\boldsymbol{\Lambda}}[v]^{\dagger} \mathcal{F}_v \left\{ \bar{\boldsymbol{r}}[m] \, \bar{\odot} \, \bar{\varphi} \left( \bar{\boldsymbol{u}}[m] \right) \right\}.
\end{equation}

\noindent where $\mathcal{F}_v(.) = \sum_{m=0}^{V-1} (.) \exp \left(-j 2 \pi \frac{mv}{V}\right) $ and $\bar{\boldsymbol{\Lambda}}[v]^{\dagger} = \left( \bar{\boldsymbol{\Lambda}}[v]^H \bar{\boldsymbol{\Lambda}}[v] \right)^{-1} \bar{\boldsymbol{\Lambda}}[v]^H$. To avoid matrix inversion, we introduce one last approximation using the channel hardening property of massive MIMO systems. Notice that $\bar{\boldsymbol{\Lambda}}[v]^{\dagger}$ from (\ref{subeqn:step1_sc}) is the zero-forcing (ZF) filter. By exploiting the large $N$, we assume $\bar{\boldsymbol{\Lambda}}[v]^H \bar{\boldsymbol{\Lambda}}[v]$ is diagonally-dominant, and $\left( \bar{\boldsymbol{\Lambda}}[v]^H \bar{\boldsymbol{\Lambda}}[v] \right)^{-1}$ can be approximated by only its diagonal part as $\boldsymbol{\lambda}[v]$ such that $\lambda_k[v] = \sfrac{1}{\sum_{n=1}^{N} | [\bar{\boldsymbol{\Lambda}}[v]]_{(n,k)}|^2 }$ for $k=1,2,\hdots,K$ and $v=0,1,\hdots,V-1$ due to the channel hardening property of massive MIMO using the LLN. Hence, we replace the ZF filter expression during the step calculations with the maximum ratio combining (MRC) filter by relying on the channel hardening property to obtain the final form

\begin{equation}
    \label{subeqn:step2_sc}
    \Delta \bar{\boldsymbol{x}}[v] \cong \frac{1}{\gamma} \sqrt{\frac{N_0}{2}} \, \boldsymbol{\lambda}[v] \odot \left( \bar{\boldsymbol{\Lambda}}[v]^H \mathcal{F}_v \left\{ \bar{\boldsymbol{r}}[m] \, \bar{\odot} \, \bar{\varphi} \left( \bar{\boldsymbol{u}}[m] \right) \right\} \right).
\end{equation}

Before starting the iterative updates, an initial solution should be found, preferably close to the optimum. The low-complexity MRC estimate \cite{OBNHD} is a suitable choice, which can be found as

\begin{equation}
    \label{eqn:initial_solution}
    \bar{\boldsymbol{x}}^{(0)}[v] = \sqrt{\frac{\pi \sigma_{y}^2}{4}} \, \boldsymbol{\lambda}[v] \odot \left( \bar{\boldsymbol{\Lambda}}[v]^H \mathcal{F}_v \left\{ \bar{\boldsymbol{r}}[m] \right\} \right)
\end{equation}

\noindent for $v=0,1,\hdots, V-1$, where $\sigma_y^2=K+N_0+\sigma_{\tau}^2$ and $\sigma_{\tau}^2$ is the variance of the quantization thresholds. The projection function $\mathcal{P}^{(t)}(.)$ can be expressed as

\begin{equation}
    \label{eqn:projection_func}
    \mathcal{P}^{(t)} \left( \underline{\bar{\boldsymbol{x}}} \right) = \left\{
    \begin{array}{ll}
    \mathcal{P}_{\mathrm{box}} \left( \underline{\bar{\boldsymbol{x}}} \right), & \quad 1 \leq t < T \\
    \mathcal{P}_{\mathrm{norm}} \left( \underline{\bar{\boldsymbol{x}}} \right), & \quad t=T,
    \end{array}
    \right.
\end{equation}

\noindent where $T$ denotes the total number of iterations. $\mathcal{P}_{\mathrm{box}}(\underline{\bar{\boldsymbol{x}}}) = \mathcal{P}_{\mathrm{box}}^I \left( \Re \{\underline{\bar{\boldsymbol{x}}}\} \right) + j \mathcal{P}_{\mathrm{box}}^Q \left( \Im \{\underline{\bar{\boldsymbol{x}}}\} \right)$, where $\mathcal{P}_{\mathrm{box}}^I(.) = \mathcal{P}_{\mathrm{box}}^Q(.)$ and it is defined as $\mathcal{P}_{\mathrm{box}}^I (.) = \sign(.) \odot \min\{|.|, M_b\}$, and each function is applied element-wise on its arguments. We define the norm projection function as $\mathcal{P}_{\mathrm{norm}}(\underline{\bar{\boldsymbol{x}}}) = \frac{\sqrt{KV}}{\lVert \underline{\bar{\boldsymbol{x}}} \rVert} \underline{\bar{\boldsymbol{x}}}$ for which we utilize the practical values of $KV$ being large with the law of large numbers (LLN) at the final iteration to scale the output. At high SNR, to avoid singular Hessian matrices, we define a damping factor as in \cite{OBNHD} such that $\zeta = \max\{1,\rho/\rho_d\}$, where $\rho_d$ is the damping SNR, and the actual $N_0$ is multiplied with $\zeta$. After the iterative updates, symbol-by-symbol minimum distance mapping is applied to the estimate for an uncoded scenario. A summary of PQND is given in Algorithm \ref{alg:pqnd}.

\begin{algorithm}[t]
    \caption{Projected Quasi-Newton Detector (PQND)}
    \setstretch{1}
    \begin{algorithmic}[1]
        \label{alg:pqnd}
        \renewcommand{\algorithmicrequire}{\textbf{Input:}}
        \renewcommand{\algorithmicensure}{\textbf{Output:}}
        \REQUIRE $\{\bar{\boldsymbol{r}}[m]\}_{m=0}^{V-1}, \{\bar{\boldsymbol{\Lambda}}[v]\}_{v=0}^{V-1}, \bar{\boldsymbol{\tau}}, \alpha, T, N_0, \zeta$
        \ENSURE  $\{{\bar{\boldsymbol{x}}}[v]\}_{v=0}^{V-1}$
        \STATE Set the initial solution $\{{\bar{\boldsymbol{x}}}[v] \gets \bar{\boldsymbol{x}}^0[v]\}_{v=0}^{V-1}$ (\ref{eqn:initial_solution})
        \STATE Apply the damping factor $N_0 \gets \zeta N_0$ \label{line:damping}
        \FOR {$t = 1$ to $T$} \label{line:start}
        \STATE Calculate $\{\bar{\boldsymbol{u}}[m]\}_{m=0}^{V-1}$ using (\ref{eqn:complex_arg_vector}) \label{line:arg_cal}
        \STATE Calculate the step $\{\Delta \bar{\boldsymbol{x}}[v]\}_{v=0}^{V-1}$ (\ref{subeqn:step2_sc})
        \STATE Update $\{{\bar{\boldsymbol{x}}}[v] \gets \mathcal{P}^{(t)} \left( {\bar{\boldsymbol{x}}}[v] - \alpha \Delta \bar{\boldsymbol{x}}[v] \right) \}_{v=0}^{V-1}$ (\ref{eqn:projection_func})
        \ENDFOR
        \RETURN $\{{\bar{\boldsymbol{x}}}[v]\}_{v=0}^{V-1}$
    \end{algorithmic}
\end{algorithm}

\section{Pseudo-Random Quantization (PRQ) Scheme} \label{sec:prq_scheme}

Each pair of ADCs in the BS RF chains has a complex-valued threshold that separately represents the I/Q parts. The thresholds are generated from the Gaussian distribution, i.e., $\bar{\boldsymbol{\tau}} \sim \mathcal{CN}(\boldsymbol{0}_{N}, \sigma_{\tau}^2 \boldsymbol{I}_{N})$ as in the conventional dithering scenario. In \cite{OBNHD}, we parameterize the variance of the thresholds using empirical observations for flat-fading channels, which depend on the system parameters. Like \cite{OBNHD}, the threshold SNR value above which PRQ is utilized is determined as

\begin{equation}
\label{eqn:rho_t}
\rho_{\mathrm{t}} =
0.15 K^2 + L_s - 2.50 \log_2(N) + 14 \textrm{ dB}
\end{equation}

\noindent where $L_s$ is the number of strong, i.e., high-power channel taps, which is determined by checking the numbers of taps with an average power stronger than that of a uniform PDP channel with the same delay spread. $L_s=3$ for the SDS and $L_s=7$ for the LDS channel models. As in \cite{OBNHD}, the variance of the quantization thresholds is found accordingly using $\sigma_{\tau}^2 = \R \left(\sfrac{E_s}{\rho_{\mathrm{t}}} - N_0\right) = \R \left( \sfrac{1}{\rho_{\mathrm{t}}} - N_0 \right)$, where $\R(x) = \max\{0,x\}$ is the unit ramp function. Hence, the threshold variance is gradually increased above the threshold SNR value. The thresholds do not need updating since the distribution of the thresholds matters rather than their different realizations due to the large values of $N$. The channel capacity can be achieved by optimizing over $\underline{\boldsymbol{x}}$ and $\underline{\boldsymbol{\tau}}$. However, jointly designing the threshold values and a codebook would not be practical in a wireless system for different channel realizations and large $N$.

\section{Computational Complexity Analysis} \label{sec:complexity}

Since $N \gg K$ for massive MIMO, the complexity-dominant parts of the algorithm are computations of $\bar{\underline{\boldsymbol{u}}}$ and $\Delta\bar{\underline{\boldsymbol{x}}}$ both with complexity $\mathcal{O}(TVN\log_2(V))+\mathcal{O}(TVNK)$. The complexities of the proposed detector and the existing works from the literature are shown in Table \ref{tab:pqnd_complexity}. The generalized expectation consistent signal recovery (GEC-SR) method \cite{BayesOptimal_OFDM} requires a quadratic complexity increase in the number of BS antennas. EM and GAMP methods, explained in detail in \cite{EM_OFDM}, require a quadratic complexity increase in the data block length. Inexact EM (IEM) \cite{IEM_OFDM} and 1BOX \cite{1BOX} methods have comparable complexity with the proposed PQND method. However, the performance of the IEM method is very similar to 1BOX, and the number of iterations for the algorithm to converge is very large \cite{IEM_OFDM}. The benchmark algorithm to compare with the proposed PQND is the 1BOX detector from \cite{1BOX}, which utilizes FDE tools with first-order optimization.

\begin{table}[t]
\centering
\caption{Complexity Comparison of PQND with Existing Methods}
\label{tab:pqnd_complexity}
\resizebox{0.9\columnwidth}{!}{%
\begin{tabular}{|c|c|}
\hline
\textbf{Detector}   & \textbf{Number of Flops}                      \\ \hline
GEC-SR \cite{BayesOptimal_OFDM} & $\mathcal{O}(TNV\log_2(NV)) + \mathcal{O}(TN^2V)$ \\ \hline
EM and GAMP \cite{EM_OFDM}   & $\mathcal{O}(TNKV^2)$                         \\ \hline
PQND, 1BOX \cite{1BOX}, and IEM \cite{IEM_OFDM} & $\mathcal{O}(TVN\log_2(V))+\mathcal{O}(TVNK)$ \\ \hline
NM                  & $\mathcal{O}(TVNK^2V^3)$                      \\ \hline
\end{tabular}%
}
\end{table}

\section{Simulation Results} \label{sec:sim_results}

\begin{figure*}[t]
\subfloat[Log-Likelihood Convergence]{\label{fig:conv_lld} \includegraphics[width=0.28\linewidth]{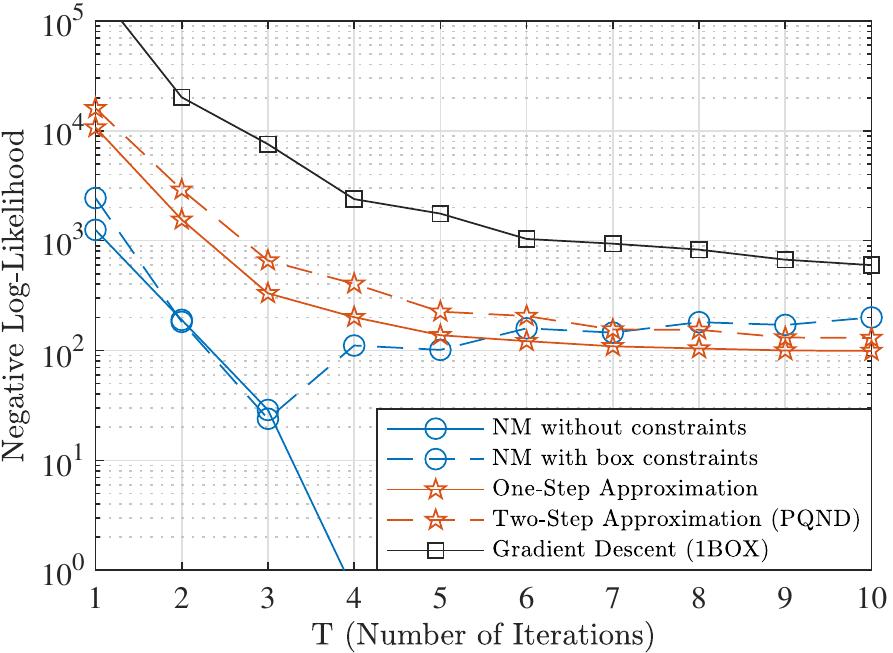}}%
\hfill
\subfloat[BER Convergence]{\label{fig:conv_ber} \includegraphics[width=0.28\linewidth]{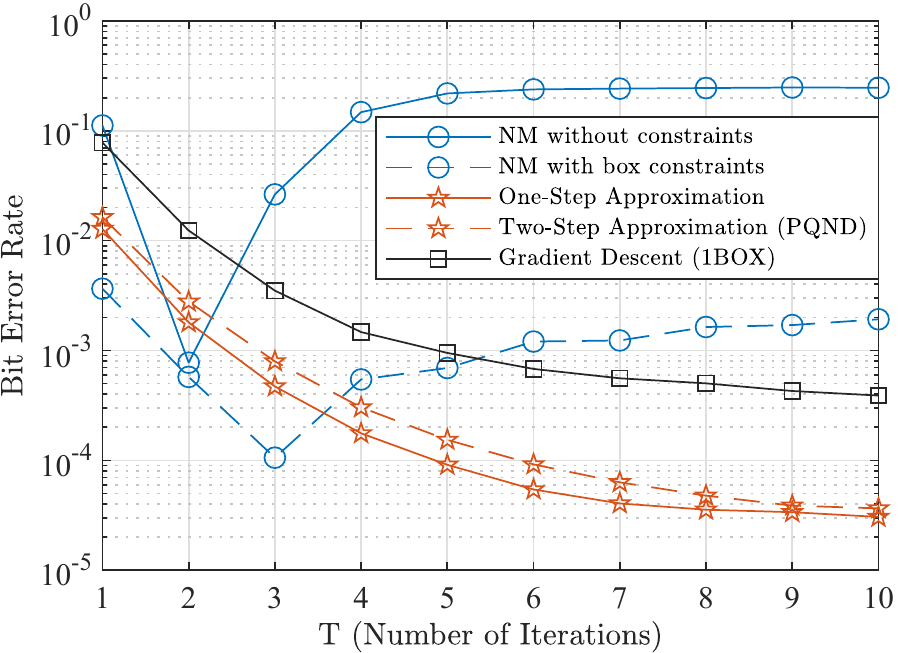}}%
\hfill
\subfloat[Performance Comparison]{\label{fig:det_comp} \includegraphics[width=0.28\linewidth]{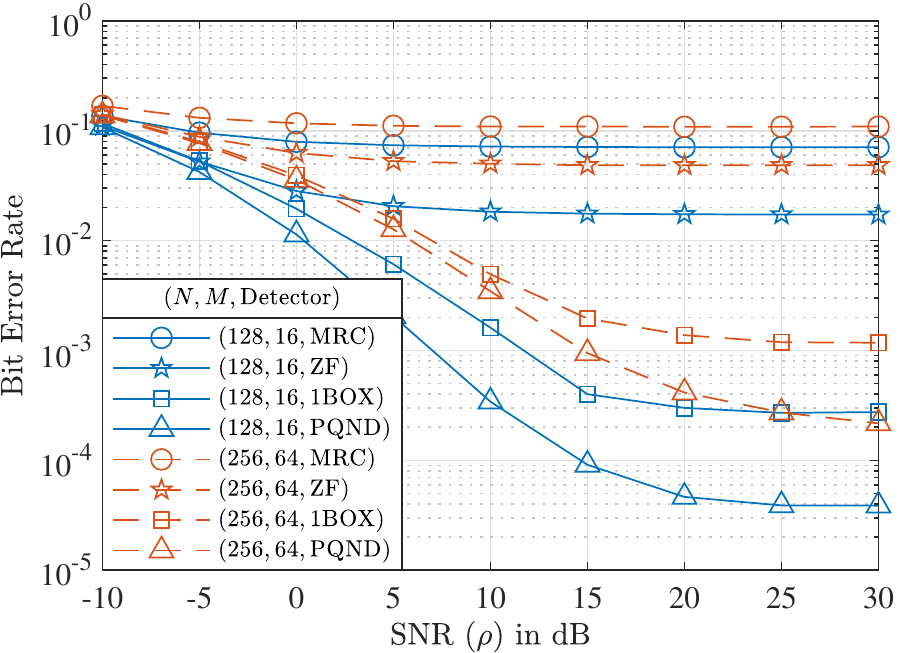}}
\caption{(a)-(b) Convergence analysis of the methods in terms of the negative of the log-likelihood and BER analysis in a $128 \times 10$ system with $V=32$, employing ZTQ and $16$-QAM constellation in the SDS channel. $\alpha=1,0.8,0.7,0.009$ for NM, its one and two-stage approximated versions, 1BOX, respectively. (c) Comparison of the BER performances of MRC, ZF, 1BOX \cite{1BOX}, and PQND methods with ZTQ in the LDS channel with respect to SNR.} \label{fig:plot1}
\end{figure*} 

In this section, the error performance of the proposed detector is investigated. PQND's step size is $\alpha=0.7$, and $T=6$ iterations are considered for all scenarios. The damping SNR for PQND is determined as $\rho_d = 20 - 150 \frac{K}{N}$ dB. The number of subcarriers is chosen as $V=256$ unless otherwise stated. An empirical convergence analysis is shown in Fig. \ref{fig:conv_lld} and \ref{fig:conv_ber}. The initial solution is the MRC estimate, and the same projection functions are used for all methods. The PQND method performs very closely to the one-step approximated method, which is a rational trade-off to avoid matrix inversion. Even after careful adjustments of the step size for the 1BOX method, it fails to converge within $10$ iterations. Unconstrained NM converges quickly in the log-likelihood sense. However, it does not converge to the optimal solution since the input comes from a discrete alphabet. Even though NM with box constraints can yield better performance for up to 3 iterations, its complexity is $O(TVNK^2 V^3)$. On the other hand, the proposed method has complexity $\mathcal{O}(TVNK)+\mathcal{O}(TVNlog_2(V))$ since it does not require matrix inversion and can utilize the fast Fourier transform (FFT) algorithm. The BER performances of different detectors from the literature are compared in Fig. \ref{fig:det_comp}. The ZF estimate can be found as $\tilde{\bar{\boldsymbol{x}}}[v] = \sqrt{\frac{\pi \sigma_{y}^2}{4}} \, \bar{\boldsymbol{\Lambda}}[v]^{\dagger} \mathcal{F}_v \left\{ \bar{\boldsymbol{r}}[m] \right\}$. $\rho_d=15$ dB, $T=6$, and $0.007$ step size are selected for 1BOX. MRC and ZF perform poorly compared to the nonlinear methods. The proposed PQND outperforms 1BOX as expected since second-order techniques converge faster.

\begin{figure*}[t]
\subfloat[PRQ vs. ZTQ]{\label{fig:two_user} \includegraphics[width=0.28\linewidth]{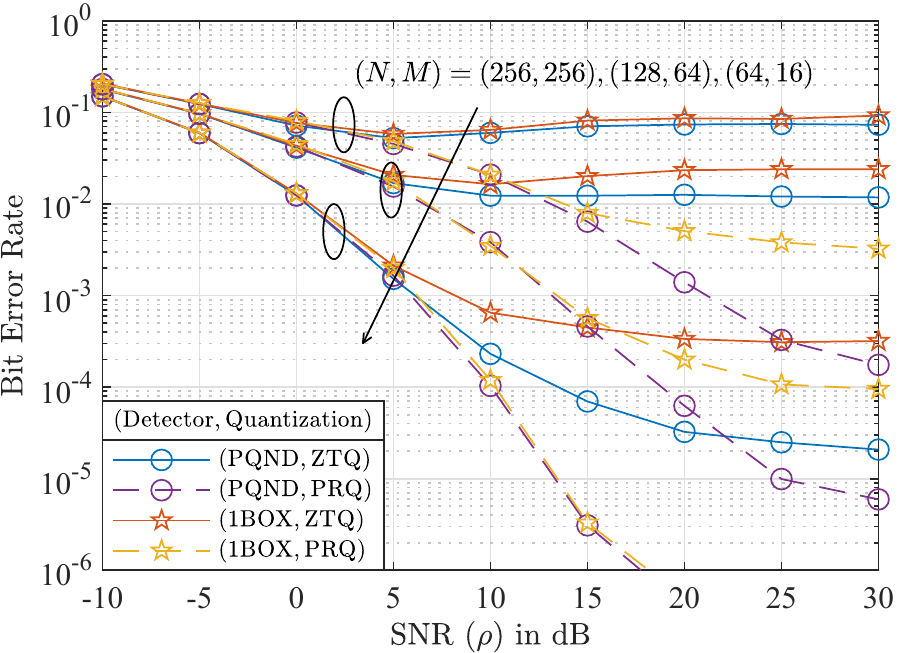}}%
\hfill
\subfloat[BER vs. $K$]{\label{fig:wrt_k} \includegraphics[width=0.28\linewidth]{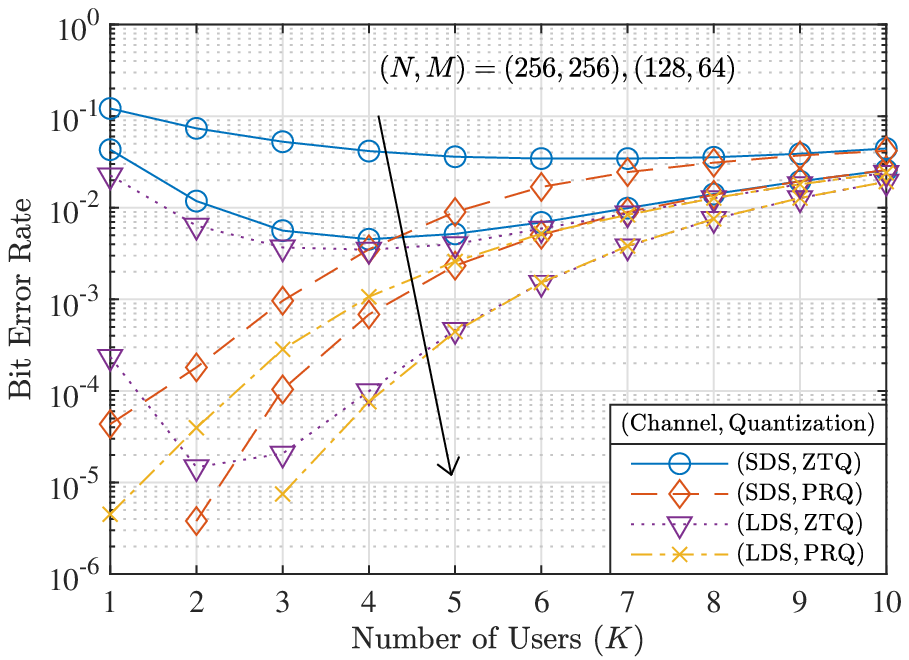}}%
\hfill
\subfloat[BER vs. $N$]{\label{fig:wrt_n} \includegraphics[width=0.28\linewidth]{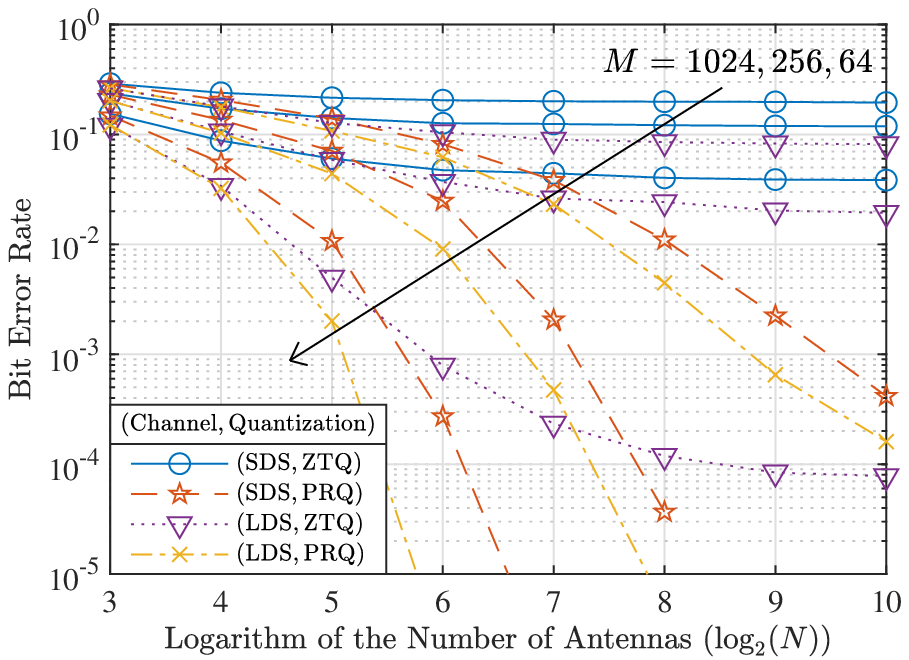}}
\caption{(a) The performance comparison of PQND and 1BOX \cite{1BOX} with respect to SNR in the SDS channel using ZTQ and PRQ where $K=2$. With 1BOX, $\alpha=0.02,0.01,0.007$ for $N=64,128,256$, respectively. (b) The BER performance against $K$ at $\rho=30$ dB with ZTQ and PRQ in the SDS and LDS channels. (c) The BER performance against $\log_2(N)$ when $K=1$ at $\rho=30$ dB with ZTQ and PRQ in the SDS and LDS channels.} \label{fig:plot2}
\end{figure*} 

In Fig. \ref{fig:two_user}, the performance is investigated when ZTQ and PRQ schemes are utilized with high-order modulations using PQND and 1BOX. The high SNR performance generally starts to saturate with ZTQ after some point. The spatial degrees of freedom are exploited much better with PRQ, and high SNR error floors are decreased to much lower levels. Existing methods for one-bit massive MIMO focus on QPSK and 16-QAM constellations. According to the results, high-order modulation schemes can also be applicable in one-bit massive MIMO-OFDM systems with the PRQ scheme. PQND and 1BOX perform similarly with PRQ for $16$-QAM, yet PQND is significantly better at higher $M$. In Fig. \ref{fig:wrt_k}, the PQND performances with respect to $K$ are plotted using ZTQ and PRQ. Since the high SNR performance is the limiting factor, $\rho=30$ dB is selected. As in \cite{OBNHD}, starting with the single-user scenario, increasing $K$ results in better performance at the beginning, which are implications of the SR phenomenon and multi-user interference serving as a source of dither. The performance gain obtained with PRQ is more significant for small $K$ values. The LDS channel exhibits a higher ISI level than the SDS channel, which leads to better performance. These results suggest that at high SNR, up to a certain point of sum interference composed of MUI and ISI, massive MIMO systems can benefit from PRQ to achieve higher rates per user by employing higher $M$. In Fig. \ref{fig:wrt_n}, we obtain the PQND performances when $K=1$ with respect to $\log_2(N)$ at $\rho=30$ dB. The SDS channel does not allow the usage of these high-order modulation schemes with the conventional ZTQ, even when $N$ is very large. Hence, ISI helps lower the amplitude distortion as suggested in \cite{Wideband_1bit_Perf}. However, even in the LDS channel, utilizing PRQ is necessary to work with $256$-QAM and $1024$-QAM.

\section{Conclusion} \label{sec:conc}

This letter proposes a new detection method called PQND that operates with PRQ for one-bit massive MIMO-OFDM systems. PQND is derived based on NM with additional approximations to obtain a quasi-Newton method by operating at the subcarrier level in the FD. By utilizing PQND and PRQ, one-bit massive MIMO-OFDM systems can support high-order modulation schemes and benefit from higher rates per user at high SNR, especially for less dispersive channels, e.g., mmWave channel. The proposed detector outperforms the benchmark method 1BOX with comparable complexity.


\bibliographystyle{IEEEtran}
\balance
\bibliography{main}

\vfill

\end{document}